# Seebeck-driven transverse thermoelectric generation in on-chip devices

Weinan Zhou[1], Takamasa Hirai[2], Ken-ichi Uchida[2,3,4] and Yuya Sakuraba[2]

[1] International Center for Young Scientists, National Institute for Materials Science,
Tsukuba 305-0047, Japan
[2] Research Center for Magnetic and Spintronic Materials, National Institute for Materials Science,
Tsukuba 305-0047, Japan
[3] Institute for Materials Research, Tohoku University, Sendai 980-8577, Japan
[4] Center for Spintronics Research Network, Tohoku University, Sendai 980-8577, Japan

E-mail: ZHOU.Weinan@nims.go.jp

## Abstract

An unconventional approach to enhance the transverse thermopower by combining magnetic and thermoelectric materials, namely the Seebeck-driven transverse thermoelectric generation (STTG), has been proposed and demonstrated recently. Here, we improve on the previously used sample structure and achieve large transverse thermopower over 40 μV K$^{-1}$ due to STTG in on-chip devices. We deposited polycrystalline Fe-Ga alloy films directly on n-type Si substrates, where Fe-Ga and Si serve as the magnetic and thermoelectric materials, respectively. Using microfabrication, contact holes were created through the SiO$_x$ layer at the top of Si to electrically connect the Fe-Ga film with the Si substrate. These thin devices with simple structure clearly exhibited enhancement of transverse thermopower due to STTG, and the obtained values agreed well with the estimation over a wide range of the size ratio between the Fe-Ga film and the Si substrate.

## 1. Introduction

Transverse thermoelectric generation—the conversion of a heat flow into a transverse charge current—has been attracting ever-growing interest in recent years. Different from the traditional thermoelectric generation based on the Seebeck effect (SE), here the direction of the generated thermopower is perpendicular to the applied temperature gradient (∇T). When converting thermal energy from a heated surface, this orthogonal relationship allows the device structure to be in a two-dimensional form made of a simple sheet or connecting wires on the surface, and the output can be enhanced by enlarging the device in plane. Comparing to the complicated three-dimensional structure made of serially connected Π-shaped junctions to exploit the SE, the simple in-plane structure using transverse thermoelectric generation is advantageous in achieving low module cost as well as high durability and flexibility [1-5]. One of the well-known transverse thermoelectric phenomena is the anomalous Nernst effect (ANE) observed in magnetic materials [6-38]. In addition to exploit it for thermoelectric generation, there has been effort to explore other functionalities, such as heat flux sensing [22,29].

Meanwhile, the application of transverse thermoelectric generation has also been stimulated by the large anomalous Nernst thermopower ($S_{ANE}$) observed in recently discovered

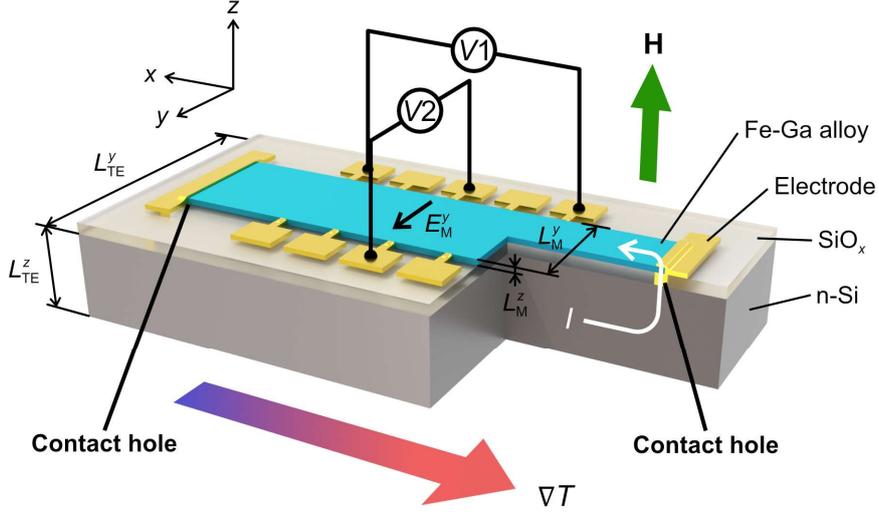

**Figure 1.** Schematic illustration of the structure and measurement set-up to demonstrate STTG in an on-chip device. The n-Si substrate is used as the thermoelectric material, while the Fe-Ga alloy film deposited on top is used as the magnetic material. The right bottom corner of the sample is removed to show its cross-section, where a contact hole is fabricated through the insulating $SiO_x$ layer. The magnetic and thermoelectric materials are electrically connected at both ends along the temperature gradient $\nabla T$ direction by the contact holes to form a closed circuit. When $\nabla T$ is applied, the charge current $I$ due to the SE of the thermoelectric material is applied to the magnetic material, and then converted to the transverse electric field $E_M^y$ by the AHE of the magnetic material. $V1$ and $V2$ represent two nanovoltmeters that measure the voltage due to the longitudinal and transverse thermopower, respectively.

magnetic topological materials. $S_{ANE}$ can be separated into two components as

$$S_{ANE} = \rho_{xx}\alpha_{xy} - \rho_{AHE}\alpha_{xx}, \quad (1)$$

where $\rho_{xx}$, $\rho_{AHE}$, $\alpha_{xx}$, and $\alpha_{xy}$ are the longitudinal resistivity, anomalous Hall resistivity, longitudinal thermoelectric conductivity, and transverse thermoelectric conductivity, respectively. The first term on the right-hand side of equation (1), $\rho_{xx}\alpha_{xy}$ (defined as $S_I$), is usually considered the intrinsic contribution to ANE since $\alpha_{xy}$ directly converts $\nabla T$ into a transverse electrical current as $j_y = \alpha_{xy}\nabla T$. It is found that large Berry curvature originated from the topological electronic structure near the Fermi level leads to large values of $\alpha_{xy}$, and usually results in the observation of large $S_{ANE}$ as well. Indeed, record-high $S_{ANE}$ values of 23 μV K$^{-1}$ in UCo$_{0.8}$Ru$_{0.2}$Al at low temperature [36], or 5–8 μV K$^{-1}$ in $L2_1$-Co$_2$MnGa and Co$_2$Mn(Al,Si) at room temperature have been reported in such topological materials [14,16,23,25,27]. In additional, magnetic materials with other properties that are favored in applications, e.g., small magnetization [11,13,18,29,34,38], negative value of $S_{ANE}$ [7,30], or large magnetic anisotropy [30,34,39] have also received attention to their transverse thermoelectric phenomena. Although orders of magnitude enhancement of $S_{ANE}$ has been made in recent year from that of the traditional magnetic materials like Fe, Co, and Ni, the thermopower is still small compared to that of the SE and further enhancement is strongly required for practical applications.

Recently, a different approach to achieve large transverse thermopower has been proposed and demonstrated by combining magnetic and thermoelectric materials [40,41]. The second term on the right-hand side of equation (1), $-\rho_{AHE}\alpha_{xx}$ (defined as $S_{II}$), is due to the anomalous Hall effect (AHE) acting on the longitudinal charge current induced by the SE, and can be rewritten as $-S_{SE}\times\rho_{AHE}/\rho_{xx} = -S_{SE}\times\tan\theta_{AHE}$, where $S_{SE}$ is the Seebeck coefficient and $\theta_{AHE}$ is the anomalous Hall angle. This term is usually insignificant for the metallic magnetic materials having small $S_{SE}$. However, large $S_{SE}$ can be found in thermoelectric materials. Therefore, one can imagine a hybrid system consisting of a magnetic material and a thermoelectric material connected at both ends along the direction of $\nabla T$ to form a closed circuit. Here, the $S_{II}$ term is artificially engineered; the large $S_{SE}$ of the thermoelectric material would induce a longitudinal charge current in the magnetic material, which is then converted to the transverse direction by the AHE of the magnetic material. This is referred to as the Seebeck-driven transverse thermoelectric generation (STTG). Based on a phenomenological calculation, the total transverse thermopower of the system, $S_{tot}^y$, is expressed as

$$S_{tot}^y = \left(\frac{E_M^y}{-\nabla T}\right) = S_{ANE} - \frac{\rho_{AHE}}{\rho_{TE}/r+\rho_M}(S_{TE} - S_M), \quad (2)$$

where $E_M^y$ is the transverse electric field along the $y$ direction in the magnetic material, $\rho_{TE(M)}$ and $S_{TE(M)}$ are the longitudinal resistivity and the Seebeck coefficient of the thermoelectric (magnetic) material, respectively. $r = (L_M^x/L_{TE}^x) \times (L_{TE}^y L_{TE}^z / L_M^y L_M^z)$ is the size ratio between the thermoelectric and magnetic materials with $L_{TE(M)}^x$, $L_{TE(M)}^y$, and $L_{TE(M)}^z$ respectively being the size of the thermoelectric (magnetic) material along the $x$, $y$, and $z$ direction. In addition to the ANE contribution from the magnetic material, the second term on the right-hand side of equation (2) describes the STTG contribution, which can reach the order of 100 µV K$^{-1}$ with the proper thermoelectric and magnetic materials and an optimized $r$. It is worth mentioning that the STTG contribution has an upper limit of tan$\theta_{AHE}$ of the magnetic material times $S_{SE}$ of the thermoelectric material. Since currently tan$\theta_{AHE}$ is still much less than 1, $S_{tot}^y$ due to STTG would be smaller than $S_{SE}$. However, due to the orthogonal relationship between $\nabla T$ and the generated thermopower, the STTG has the aforementioned merits of transverse thermoelectric generation when applied to thermoelectric modules. The STTG has been verified through an experimental demonstration [40], where the sample consisted of an epitaxial $L2_1$-Co$_2$MnGa film, deposited on a single-crystal MgO substrate, and a Si substrate. When Co$_2$MnGa showing the large AHE and n(p)-type Si showing the large SE were connected through bonding wires to form a closed circuit, the hybrid structure exhibited a large value of $S_{tot}^y$ = 82.3 (−41.0) µV K$^{-1}$, which also agreed quantitatively with the estimation using equation (2). Although this first demonstration of STTG observed the highest transverse thermopower ever reported while preserving the most important advantage of transverse thermoelectric generation, *i.e.*, the orthogonal relationship between $\nabla T$ and the electric field, the fabricated device structure was complex to be applied for any application; the binding of two substrates made the sample much thicker than necessary, and bonding wires complicated the structure. Therefore, improvement on the device structure is strongly required in order to further explore the potential of STTG.

In this study, we demonstrate the STTG in on-chip devices having a simplified structure, as shown in figure 1. The thermally oxidized n-type Si (n-Si) substrate is used as the thermoelectric material, while the disordered Fe-Ga alloy film deposited on top functions as the magnetic material. The Fe-Ga alloy was chosen because it can be deposited easily by sputtering while showing a considerably large AHE. A previous study has shown strong enhancement of AHE and ANE in epitaxial Fe-Ga alloy films compared to that of Fe, and attributed this to the Fermi-level tuning through substitution of Fe atoms with Ga atoms [20]. Here, we prepared and evaluated polycrystalline Fe-Ga alloy films having thickness of 10–100 nm, and found similarly large AHE and ANE, with small thickness dependence compared to that of $L2_1$-Co$_2$MnGa [42]. Microfabrication processes were used to create contact holes between the Fe-Ga film and the n-Si to form a closed circuit, resulting in a thin and simple structure. Together with the control of the shape of Fe-Ga alloy films through microfabrication, we were able to change $L_M^y$ and $L_M^z$ to achieve an approximately two orders of magnitude variation in $r$ among different samples. The measurement results clearly exhibited the STTG contribution, with $S_{tot}^y$ of > 40 µV K$^{-1}$ in samples with large $r$. The results also agreed well with the estimation using equation (2), validating the feasibility of the much simple device structure without the binding of two substrates and bonding wires.

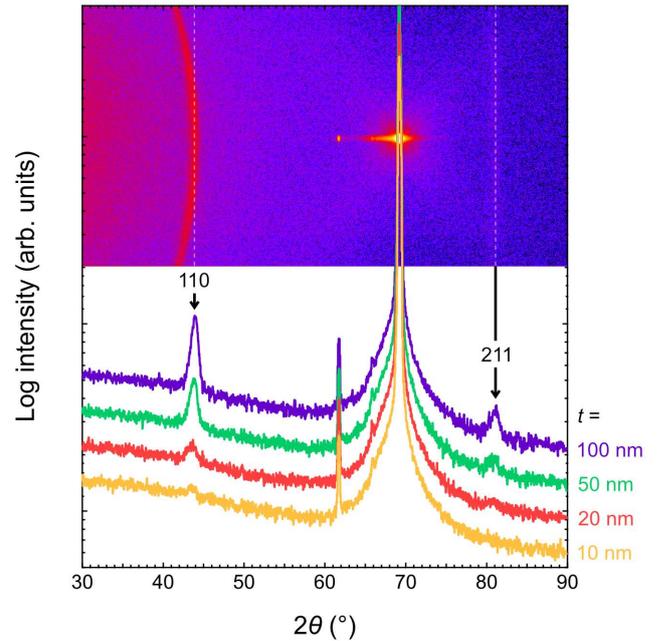

**Figure 2.** Out-of-plane XRD patterns of the Fe-Ga alloy blanket films with the thickness $t$ = 10, 20, 50, and 100 nm. The inset colour map at the top is the two-dimensional XRD pattern of the $t$ = 100 nm sample. The sharp peaks at $2\theta$ = 62° and 69° are the 400 diffraction of the Si substrate due to Cu-K$\beta$ and Cu-K$\alpha$, respectively.

## 2. Experimental procedures

The on-chip devices were prepared following a bottom-up procedure. The phosphorous-doped n-Si substrate was cut into the size of 5 mm in width and 10 mm in length. Its thickness was 0.525 mm, with a 100-nm-thick SiO$_x$ insulator layer at the top surface. First, the contact holes were fabricated. Using photolithography and Ar-ion milling, we etched away two patches of the SiO$_x$ layer with the size of 0.2 mm in the $x$ direction, 3 mm in the $y$ direction, and 8 mm spacing between their centers (see figure 1). After reaching the n-Si, these two contact holes were filled with Ta (2 nm) / Au (100 nm) films deposited by sputtering without breaking the vacuum. Then, using a lift-off process, the Fe-Ga alloy films were deposited from a single Fe$_{65}$Ga$_{35}$ alloy target by magnetron sputtering at room temperature. After the patterning of the photoresist and prior to the deposition, Ar-ion milling was used to clean the surface of the samples. The thickness of the Fe-Ga alloy films ($t$) was set at 10, 20, 50, and 100 nm. 1-nm-thick Au capping layers were deposited on top of the Fe-Ga alloy films to prevent oxidation. Finally, the Au electrode pads for transport measurements were fabricated using a lift-off process. The size of the Fe-Ga alloy film in the $y$ direction was set at 2 mm, except for the $t$ = 20 nm samples, where the devices having the sizes of 1, 0.5, and 0.25 mm in the $y$ direction were also prepared. During the fabrication of the devices, we paid special attention to prevent metallic films deposited on the side of the n-Si substrate, which would shunt the charge current generated by the SE of n-Si. Reference samples without the contact holes, as well as blanket films, were also prepared in order to investigate the properties of the Fe-Ga alloy films.

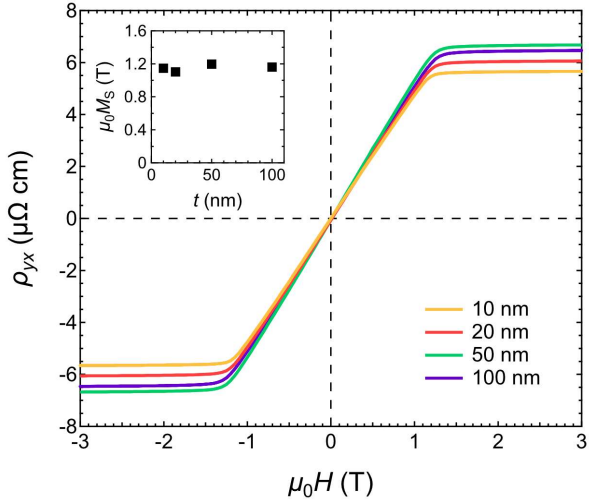

**Figure 3.** Magnetic field $H$ dependence of the transverse resistivity $\rho_{yx}$ for the reference Fe-Ga alloy film samples with $t$ = 10, 20, 50, and 100 nm. The inset shows the $t$ dependence of the saturation magnetization $M_s$ of the Fe-Ga alloy films.

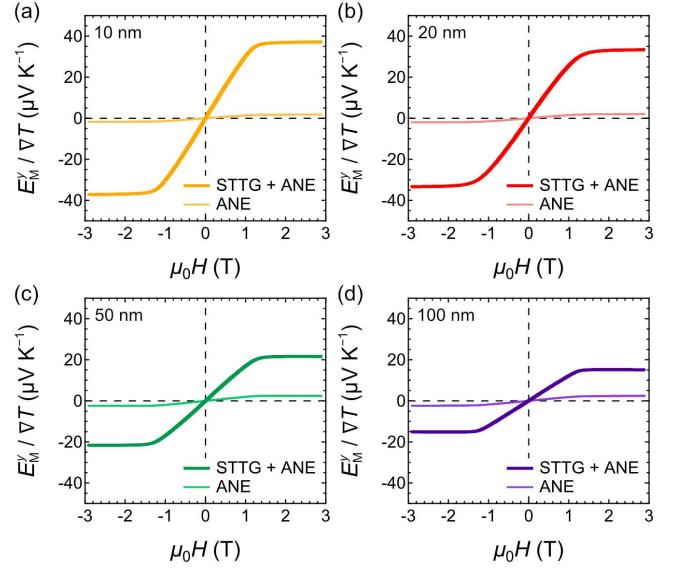

**Figure 4.** $H$ dependence of the transverse electric field $E_M^y$ divided by $\nabla T$ for the on-chip device (STTG + ANE) and the reference sample (ANE) with $t$ = (a) 10, (b) 20, (c) 50, and (d) 100 nm.

The composition of the Fe-Ga alloy films was determined to be Fe$_{72}$Ga$_{28}$ by wavelength dispersive X-ray fluorescence analysis. The structure of the films was studied using an X-ray diffraction (XRD) equipped with Cu-$K\alpha$ radiation and a two-dimensional detector. The magnetic properties were measured with a vibrating sample magnetometer. $\rho_M$ and $\rho_{AHE}$ of the Fe-Ga alloy films were measured from reference samples using a physical property measurement system (PPMS, Quantum Design). $S_M$ and $S_{ANE}$ of the Fe-Ga alloy films as well as $S_{tot}^y$ of the on-chip devices were measured using a home-made holder embedded in a multi-function probe, which was used with the PPMS. An infrared camera was used to measure the temperature of the samples, which was done outside the PPMS.

The surfaces of the samples were partially covered with black ink having a known emissivity of 0.94 for the temperature measurement. When a certain charge current was applied to the Peltier module in the home-made holder, a stable $\nabla T$ across the sample can be generated after the holder reaches thermal equilibrium. The voltage due to the longitudinal thermopower was measured ($V1$, see figure 1), while the temperatures of the sample at the positions corresponding to the two contacts to $V1$ along the direction of $\nabla T$ was extract from the infrared camera images. Using the on-chip device having 10-nm-thick Fe-Ga film as an example, when we applied 1.0, 0.8, −0.8, and −1.0 A to the Peltier module, we obtained the temperature difference between the two positions to be 4.42±0.16, 3.55±0.16, −3.63±0.16, and −4.58±0.16 K, respectively. With the known distance of 4 mm between these two positions, $\nabla T$ was calculated, which showed a linear relationship with the voltage from $V1$. Then the home-made holder was placed inside the PPMS to carry out the measurement while $H$ along the $z$ direction was varied. The voltage of $V1$ under zero $H$ was used to estimate $\nabla T$ based on the previously obtained linear relationship. Together with the voltage of $V2$, $S_{tot}^y$ of the on-chip devices (or $S_{ANE}$ of the Fe-Ga films) can be obtained (see details of the measurement in [40]). All the measurements were performed at room temperature.

## 3. Results and discussion

The structure of the Fe-Ga alloy films was studied by measuring the out-of-plane XRD patterns of the blanket films, and the results are shown in figure 2. The 110 diffraction peak of the Fe-Ga alloy can be seen from all samples, with thicker samples also showing the 211 peak. The intensity of the 110 and 211 peaks increases with increasing $t$ mainly due to larger volume of Fe-Ga diffracting the X-ray. The inset colour map is the two-dimensional XRD pattern of the sample with $t = 100$ nm. One can clearly see the ring-shaped 110 and 211 peaks, indicating that the Fe-Ga alloy film is randomly oriented polycrystalline. No peak from the ordered Fe-Ga phases was observed, suggesting the Ga atoms are substituting the Fe atoms randomly and the Fe-Ga alloy forms atomic disordered structure. This is reasonable considering the films were deposited at room temperature without thermal treatment.

The transport properties of the Fe-Ga alloy films, i.e., $\rho_M$, $\rho_{AHE}$, $S_M$, and $S_{ANE}$, were evaluated using the reference samples. $\rho_M$ was obtained using the 4-terminal method under zero external magnetic field ($H$). The $H$ dependence of the transverse resistivity ($\rho_{yx}$) is shown in figure 3. Here, $H$ was applied along the $z$ direction (see figure 1). Note that the 1-nm-thick Au capping layer was also taken into consideration for calculating the resistivity. One can see that $\rho_{yx}$ saturated with $|\mu_0 H| > 1.2$ T, and $\rho_{AHE}$ was obtained by linearly extrapolating the data points of $\rho_{yx}$ after saturation to zero $H$. The results show small difference among the samples with different $t$. The inset of figure 3 shows the $t$ dependence of the saturation magnetization ($M_s$) of the Fe-Ga alloy films, obtained from the hysteresis loops of the blanket films. A similar $\mu_0 M_s \sim 1.2$ T was obtained for the samples with different $t$, which also corresponds well with the values of $\mu_0 H$ for $\rho_{yx}$ to saturate.

To evaluate the transverse thermopower, we measured the electrical output along the $y$ direction while sweeping $H$. The measured curves of $E_M^y$ divided by $\nabla T$ as a function $H$ are shown in figure 4 for the reference samples (narrow line) as well as the on-chip devices (broad line). The values of transverse thermopower due to the spontaneous magnetization were determined in a similar manner to $\rho_{AHE}$ by linear extrapolating the data after saturation to zero $H$, in order to exclude the contribution of the ordinary Nernst effect. The output from the reference samples is due to the ANE, and $S_{ANE}$ of the Fe-Ga alloy films exhibits a slight increase with increasing $t$. $S_{ANE}$ with different $t$ together with other measured transport properties of the Fe-Ga alloy films are summarized in table 1. On the other hand, the on-chip devices showed significant enhancement of transverse thermopower due to STTG, with the highest value of $S_{tot}^y = 36.4\pm0.5$ $\mu$V K$^{-1}$ from the device of $t = 10$ nm, which is more than one order of magnitude larger than $S_{ANE}$. $S_{tot}^y$ decreases with increasing $t$, down to 15.4±0.6 $\mu$V K$^{-1}$ for the device of $t = 100$ nm. The reason of this behavior is that the increase of $L_M^z$ leads to a decrease of $r$. While $S_{tot}^y$ contains the anomalous Nernst effect of the Fe-Ga films as shown in equation (2), the contribution of STTG is the difference between the broad and narrow curves in figure 4, which can be obtained by subtracting $S_{ANE}$ from $S_{tot}^y$.

**Table 1.** Transport properties of the Fe-Ga alloy reference films with different $t$.

| $t$ (nm) | $\rho_M$ ($\mu\Omega$ cm) | $\rho_{AHE}$ ($\mu\Omega$ cm) | $S_M$ ($\mu$V K$^{-1}$) | $S_{ANE}$ ($\mu$V K$^{-1}$) |
|---|---|---|---|---|
| 10 | 126.13±0.01 | 5.60±0.01 | −14.5±0.3 | 1.65±0.04 |
| 20 | 125.90±0.01 | 5.99±0.01 | −15.9±0.5 | 1.91±0.05 |
| 50 | 132.75±0.01 | 6.60±0.01 | −17.8±1.6 | 2.35±0.14 |
| 100 | 129.64±0.01 | 6.38±0.01 | −17.9±1.8 | 2.30±0.15 |

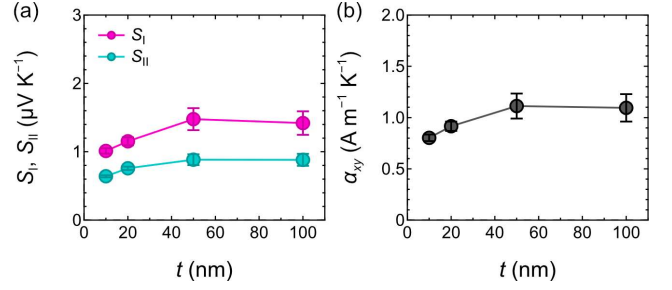

**Figure 5.** (a) $S_I$ and $S_{II}$ terms of the anomalous Nernst coefficient $S_{ANE}$, and (b) transverse thermoelectric conductivity $\alpha_{xy}$ of the Fe-Ga alloy reference films as a function of $t$.

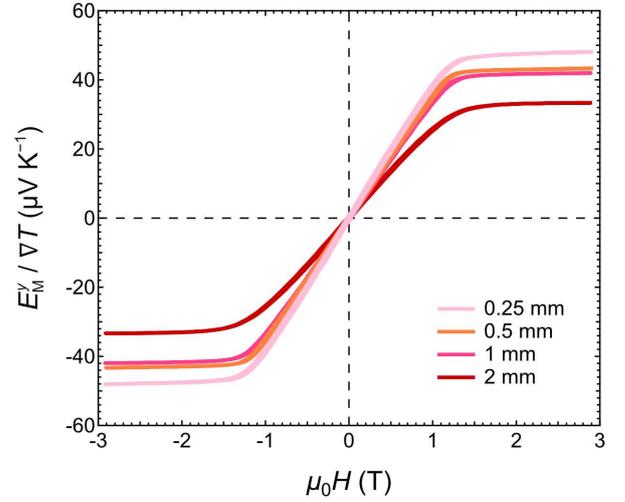

**Figure 6.** $H$ dependence of $E_M^y$ divided by $\nabla T$ for the on-chip devices having different Fe-Ga width $L_M^y$.

The transport properties in table 1 allow us to further analyze the polycrystalline Fe-Ga alloy films using equation (1). $S_{ANE}$ is separated into the $S_I$ and $S_{II}$ terms, as shown in figure 5(a). One can see that the intrinsic $S_I$ term is larger than the $S_{II}$ term. From this result, we obtained $\alpha_{xy} \sim 1$ A m$^{-1}$ K$^{-1}$ for the Fe-Ga alloy films. $\alpha_{xy}$ also shows a slight increase with increasing $t$ (see figure 5(b)), which is attributable to the decreasing influence from the Au capping layer and better crystallinity in thicker films. Overall, the properties of the polycrystalline Fe-Ga films reported here are similar to that of the epitaxial film having the same composition [20].

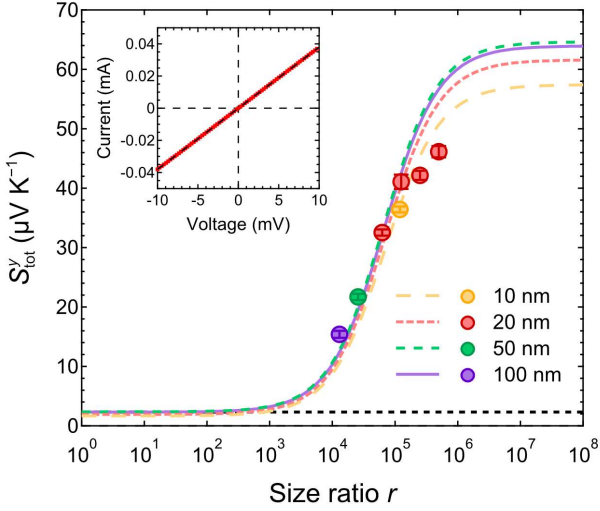

**Figure 7.** Size ratio $r$ dependence of the total transverse thermopower $S_{\text{tot}}^y$. The curves are calculated using equation (2), with the transport properties of the Fe-Ga alloy films shown in table 1 and the properties of n-Si described in the manuscript. The data points are obtained from the on-chip devices plotted at the corresponding $r$. The black dashed line at the bottom represents $S_{\text{ANE}}$ of the Fe-Ga alloy film with $t = 100$ nm. The inset shows the current-voltage curve measured from the n-Si substrate with Ta/Au contact holes. The dashed line represents a linear fit.

The microfabrication processes for the on-chip devices allow us to easily investigate the $L_M^y$ dependence of $S_{\text{tot}}^y$, which was not studied in the previous experiments [40]. For $t = 20$ nm, we prepared additional devices with $L_M^y = 1$, 0.5, and 0.25 mm. Figure 6 shows the measured curves of $E_M^y$ divided by $\nabla T$ as a function $H$ for these devices. The result from the device having $L_M^y = 2$ mm, which has been shown in figure 4(b), is also added for comparison. There is an unambiguous increase of $S_{\text{tot}}^y$ with decreasing $L_M^y$. In case of the ANE, $S_{\text{ANE}}$ is determined by the material, and the value should remain unchanged when the size of the film is varied. On the other hand, for the on-chip devices, decreasing $L_M^y$ leads to an increase of $r$ and $S_{\text{tot}}^y$.

Finally, we compare the values of $S_{\text{tot}}^y$ obtained experimentally from the on-chip devices with the estimation using equation (2), as shown in figure 7. The values of $S_{\text{tot}}^y$ measured from the on-chip devices are plotted as data points at the corresponding $r$. Here, to obtain $r$ of the devices, we assumed $L_M^x = L_{\text{TE}}^x$, $L_M^y$ of 2, 1, 0.5, and 0.25 mm, and $L_M^z$ of 11, 21, 51, and 101 nm, where the Au capping layer was taken into consideration. Because there is small difference in the transport properties of the Fe-Ga alloy films with different $t$, we calculated the $r$ dependence of $S_{\text{tot}}^y$ based on different sets of values of the transport properties from $t = 10$, 20, 50, and 100 nm (see table 1). The n-Si substrates are the same ones used in [40], having $S_{\text{TE}} = -1.3$ mV K$^{-1}$, and high thermal conductivity $\kappa_{\text{TE}} = 1.5 \times 10^2$ W m$^{-1}$ K$^{-1}$. While the current study is focusing on $S_{\text{tot}}^y$, the efficiency of thermoelectric generation using STTG requires the optimization of the materials with the thermal conductivity taken into consideration, which requires further studies. The tendency of the calculated $r$ dependence of $S_{\text{tot}}^y$ is similar, while thick Fe-Ga alloy films with large $\tan\theta_{\text{AHE}}$ lead to further enhancement of $S_{\text{tot}}^y$ when $r$ increases. Overall, $S_{\text{tot}}^y$ of the on-chip devices agrees well with the calculated $r$ dependence of $S_{\text{tot}}^y$. For the two on-chip devices having $L_M^y = 0.5$ and 0.25 mm, although the measured $S_{\text{tot}}^y$ follows the tendency of increasing with increasing $r$, the values deviate from the calculation as shown by the two data points with the largest $r$ in figure 7. The reason behind this deviation is not clear but might be due to the contact resistance. Since the decrease in $L_M^y$ reduces the contact area between the Fe-Ga alloy film and the contact holes, it may make it difficult for the charge current from n-Si to flow into Fe-Ga, and change the current distribution in the n-Si substrate. It is worth mentioning that we used an equivalent resistivity $\rho'_{\text{TE}}$ for the calculation instead of the resistivity of n-Si, in order to account for the additional resistance to the closed circuit from the Ta/Au of the contact holes, the interface between the Ta/Au and the n-Si, as well as the possible damage to the n-Si during microfabrication. We measured the current-voltage curve of the n-Si substrate with the Ta/Au contact holes, as shown in the inset of figure 7. Here, we used the 4-terminal method, by attaching the positive leads to one contact hole and the negative leads to the other contact hole. The linear relationship between the current and voltage indicates ohmic contacts to the n-Si. A linear fit yields a resistance $R'_{\text{TE}}$ of 263 Ω. Since the term containing $\rho_{\text{TE}}$ and $r$ in equation (2) can be rewritten as $\rho_{\text{TE}}/r = \rho_{\text{TE}}\left(L_{\text{TE}}^x L_M^y L_M^z / L_M^x L_{\text{TE}}^y L_{\text{TE}}^z\right) = \left(\rho_{\text{TE}} L_{\text{TE}}^x / L_{\text{TE}}^y L_{\text{TE}}^z\right) \times \left(L_M^y L_M^z / L_M^x\right) = R_{\text{TE}}\left(L_M^y L_M^z / L_M^x\right)$, and $L_{\text{TE}}^x = 8$ mm, $L_{\text{TE}}^y = 5$ mm, and $L_{\text{TE}}^z = 0.525$ mm are the same for all the on-chip devices, we derived $\rho'_{\text{TE}} = 8.6 \times 10^{-2}$ Ω m and used this value to calculate the $r$ dependence of $S_{\text{tot}}^y$ shown in figure 7. The obtained $\rho'_{\text{TE}}$ is twice the value of the resistivity of n-Si; the increase of $\rho'_{\text{TE}}$ tend to shift the $r$ dependence of $S_{\text{tot}}^y$ toward right. Nevertheless, $S_{\text{tot}}^y$ of the on-chip devices exhibits the effect of STTG and a significant enhancement from $S_{\text{ANE}}$ of the Fe-Ga alloy film indicated by the black dashed line. These results clearly demonstrate that simple structure of a single magnetic film on a thermoelectric substrate enables us to exploit STTG, which is an important step toward the applications of STTG.

## 4. Summary

Significant enhancement of transverse thermopower due to the STTG was achieved using on-chip devices. Here, the polycrystalline Fe-Ga alloy film was chosen as the magnetic material and directly deposited on the n-Si substrate that functioned as the thermoelectric material. Microfabrication processes were used to create contact holes through the insulating $SiO_x$ layer at the top surface of n-Si to form a closed circuit, leading to a thin device with simple structure. This on-chip device is a clear structural improvement from the sample used in the previous experiment, where a magnetic film deposited on a substrate was bonded to another substrate and bonding wires were used to complete the closed circuit. A series of devices having different $L_M^y$ and $L_M^z$ were prepared and evaluated. The obtained $S_{tot}^y$ agrees well with the estimated values in a wide range of $r$, reaching over 40 μV K$^{-1}$, and demonstrating the feasibility of the improved on-chip device structure. Combining magnetic and thermoelectric materials for the STTG would be a promising approach to achieve large transverse thermopower; the on-chip devices demonstrated here can serve as a platform for exploring various material combinations, and shed light on future transverse thermoelectric applications.

## Data availability statement

The data that support the findings of this study are available upon reasonable request from the authors.

## Acknowledgements

We thank S. Kasai and N. Kojima for their support in sample preparation. This work was supported by CREST "Creation of Innovative Core Technologies for Nano-enabled Thermal Management" (No. JPMJCR17I1) and Mitou challenge 2050 (No. P14004) from NEDO, Japan.

## References

[1] Bauer G E W, Saitoh E and Van Wees B J 2012 *Nat. Mater.* **11** 391
[2] Boona S R, Myers R C and Heremans J P 2014 *Energy Environ. Sci.* **7** 885
[3] Uchida K, Adachi H, Kikkawa T, Kirihara A, Ishida M, Yorozu S, Maekawa S and Saitoh E 2016 *Proc. IEEE* **104** 1946
[4] Uchida K, Zhou W and Sakuraba Y 2021 *Appl. Phys. Lett.* **118**, 140504
[5] Boona S R, Jin H and Watzman S 2021 *J. Appl. Phys.* **130** 171101
[6] Mizuguchi M, Ohata S, Uchida K, Saitoh E and Takanashi K 2012 *Appl. Phys Express* **5** 093002
[7] Sakuraba Y, Hasegawa K. Mizuguchi M, Kubota T, Mizukami S, Miyazaki T and Takanashi K 2013 *Appl. Phys Express* **6** 033003
[8] Ramos R, Aguirre M H, Anadón A, Blasco J, Lucas I, Uchida K, Algarabel P A, Morellón L, Saitoh E and Ibarra M R 2014 *Phys. Rev. B* **90** 054422
[9] Uchida K, Kikkawa T, Seki T, Oyake T, Shiomi J, Qiu Z, Takanashi K and Saitoh E 2015 *Phys. Rev. B* **92** 094414
[10] Sakuraba Y 2016 *Scr. Mater.* **111** 29
[11] Ikhlas M, Tomita T, Koretsune T, Suzuki M-T, Nishio-Hamane D, Arita R, Otani Y and Nakatsuji S 2017 *Nat. Phys.* **13** 1085
[12] Yang Z, Codecido E A, Marquez J, Zheng Y, Heremans J P and Myers R C 2017 *AIP Adv.* **7** 095017
[13] Li X, Xu L, Ding L, Wang J, Shen M, Lu X, Zhu Z and Behnia K 2017 *Phys. Rev. Lett.* **119** 056601
[14] Sakai A, Mizuta Y P, Nugroho A A, Sihombing R, Koretsune T, Suzuki M-T, Takemori N, Ishii R, Nishio-Hamane D, Arita R, Goswami P and Nakatsuji S 2018 *Nat. Phys.* **14** 1119
[15] Hu J, Ernst B, Tu S, Kuveždić M, Hamzić A, Tafra E, Basletić M, Zhang Y, Markou A, Felser C, Fert A, Zhao W, Ansermet J-P and Yu H 2018 *Phys. Rev. Appl.* **10** 044037
[16] Guin S N, Manna K, Noky J, Watzman S J, Fu C, Kumar N, Schnelle W, Shekhar C, Sun Y, Gotth J and Felser C 2019 *NPG Asia Mater.* **11** 16
[17] Guin S N, Vir P, Zhang Y, Kumar N, Watzman S J, Fu C, Liu E, Manna K, Schnelle W, Gooth J, Shekhar C, Sun Y and Felser C 2019 *Adv. Mater.* **31** 1806622
[18] Wuttke C, Caglieris F, Sykora S, Scaravaggi F, Wolter A U B, Manna K, Süss V, Shekhar C, Felser C, Büchner B and Hess C 2019 *Phys. Rev. B* **100**, 085111
[19] Ding L, Koo J, Xu L, Li X, Lu X, Zhao L, Wang Q, Yin Q, Lei H, Yan B, Zhu Z and Behnia K 2019 *Phys. Rev. X* **9** 041061
[20] Nakayama H, Masuda K, Wang J, Miura A, Uchida K, Murata M and Sakuraba Y 2019 *Phys. Rev. Mater.* **3** 114412
[21] Cox C D W, Caruana A J, Cropper M D and Morrison K 2020 *J. Phys. D: Appl. Phys.* **53** 035005
[22] Zhou W and Sakuraba Y 2020 *Appl. Phys. Express* **13** 043001
[23] Sakuraba Y, Hyodo K, Sakuma A and Mitani S 2020 *Phys. Rev. B* **101** 134407
[24] Sakai A, Minami S, Koretsune T, Chen T, Higo T, Wang Y, Nomoto T, Hirayama M, Miwa S, Nishio-Hamane D, Ishii F, Arita R and Nakatsuji S 2020 *Nature* **581**, 52
[25] Xu L, Li X, Ding L, Chen T, Sakai A, Fauqué B, Nakatsuji S, Zhu Z and Behnia K 2020 *Phys. Rev. B* **101** 180404(R)
[26] Khadka D, Thapaliya T R, Parra S H, Wen J, Need R, Kikkawa J M and Huang S X 2020 *Phys. Rev. Mater.* **4** 084203
[27] Sumida K, Sakuraba Y, Masuda K, Kono T, Kakoki M, Goto K, Zhou W, Miyamoto K, Miura Y, Okuda T and Kimura A 2020 *Commun. Mater.* **1** 89
[28] Seki T, Sakuraba Y, Masuda K, Miura A, Tsujikawa M, Uchida K, Kubota T, Miura Y, Shirai M and Takanashi K 2021 *Phys. Rev. B* **103** L020402
[29] Higo T, Li Y, Kondou K, Qu D, Ikhlas M, Uesugi R, Nishio-Hamane D, Chien C L, Otani Y and Nakatsuji S 2021 *Adv. Funct. Mater.* **31** 2008971
[30] Zhou W, Masuda K and Sakuraba Y 2021 *Appl. Phys. Lett.* **118** 152406
[31] Mende F, Noky J, Guin S N, Fecher G H, Manna K, Adler P, Schnelle W, Sun Y, Fu C and Felser C 2021 *Adv. Sci.* **8** 2100782
[32] Zhang H, Xu C Q and Ke X 2021 *Phys. Rev. B* **103** L201101

[33] Modak R, Goto K, Ueda S, Miura Y, Uchida K and Sakuraba Y 2021 *APL Mater.* **9** 031105
[34] Isogami S, Masuda K, Miura Y, Rajamanickam N and Sakuraba Y 2021 *Appl. Phys. Lett.* **118**, 092407
[35] Hamada Y, Kurokawa Y, Yamauchi T, Hanamoto H and Yuasa H 2021 *Appl. Phys. Lett.* **119** 152404
[36] Asaba T, Ivanov V, Thomas S M, Savrasov S Y, Thompson J D, Bauer E D and Ronning F 2021 *Sci. Adv.* **7** eabf1467
[37] Chen T, Minami S, Sakai A, Wang Y, Feng Z, Nomoto T, Hirayama M, Ishii R, Koretsune T, Arita R and Nakatsuji S 2022 *Sci. Adv.* **8** eabk1480
[38] Pan Y, Le C, He B, Watzman S J, Yao M, Gooth J, Heremans J P, Sun Y and Felser C 2022 *Nat. Mater.* **21** 203
[39] Miura A, Sepehri-Amin H, Masuda K, Tsuchiura H, Miura Y, Iguchi R, Sakuraba Y, Shiomi J, Hono K and Uchida K 2019 *Appl. Phys. Lett.* **115** 222403
[40] Zhou W, Yamamoto K, Miura A, Iguchi R, Miura Y, Uchida K and Sakuraba Y 2021 *Nat. Mater.* **20** 463
[41] Yamamoto K, Iguchi R, Miura A, Zhou W, Sakuraba Y, Miura Y and Uchida K 2021 *J. Appl. Phys.* **129** 223908
[42] Zhang Y, Yin Y, Dubuis G, Butler T, Medhekar N V and Granville S 2021 *npj Quantum Mater.* **6** 17